\documentclass{article}
\usepackage{spconf,amsmath,graphicx}

\usepackage[table,xcdraw]{xcolor}

\usepackage{array}
\usepackage{graphicx}
\usepackage{amsmath}
\usepackage{amssymb}
\usepackage{booktabs}

\usepackage{multirow}

\usepackage{booktabs} %

\usepackage{hyperref}

\renewcommand{\paragraph}[1]{\textbf{#1}\enspace}

\newcommand{\mosv}[2]{#1$\,\pm\,$#2}

\newcommand{\multilinecell}[3][c]{\begin{tabular}[#1]{@{}#2@{}}#3\end{tabular}}%

\newcommand{\cellv}[2]{\multilinecell{l}{#1\\#2}}

\usepackage{textcomp}
\newcommand{\textapprox}{\raisebox{0.1ex}{\texttildelow}}    

\DeclareMathOperator*{\median}{median}
  
\newcommand{\sref}[1]{\S\ref{#1}}
\newcommand{\tabref}[1]{Table~\ref{#1}}

\newcommand{\lam}{\theta}
\newcommand{\reals}{\mathbb{R}}

\newcommand{\ji}{{j(i)}}
\newcommand{\normal}{\,\mathcal{N}}

\DeclareMathOperator*{\argmax}{arg\,max}

\newcommand{\tacovb}{TacoSpawn-VB}
\newcommand{\dvector}{d-vector}

\newcommand{\demopagelink}{\url{https://google.github.io/tacotron/publications/speaker_generation}}

\newif\ifarxiv
\arxivtrue

\title{SPEAKER GENERATION}
\name{
\begin{tabular}{c}
Daisy Stanton \quad
Matt Shannon \quad
Soroosh Mariooryad \quad
RJ Skerry-Ryan \quad
\\
Eric Battenberg \quad
Tom Bagby \quad
David Kao
\end{tabular}
}
\address{Google Research, USA}

\begin{document}
\maketitle
\begin{abstract}
This work explores the task of synthesizing speech in non-existent human-sounding voices. We call this task ``speaker generation'', and present TacoSpawn, a system that performs competitively at this task. TacoSpawn is a recurrent attention-based text-to-speech model that learns a distribution over a speaker embedding space, which enables sampling of novel and diverse speakers. Our method is easy to implement, and does not require transfer learning from speaker ID systems. We present objective and subjective metrics for evaluating performance on this task, and demonstrate that our proposed objective metrics correlate with human perception of speaker similarity.  Audio samples are available on our demo page\footnote{\demopagelink}.

\end{abstract}
\section{Introduction}
\label{sec:intro}

In recent years, researchers have shown that text-to-speech (TTS) models based on deep neural networks can generate high-fidelity audio that humans often can't distinguish from genuine speech.  A major limitation of most of these models, however, is that they can only 
synthesize the voices of the human speakers used in the training dataset.  Expanding the set of voices for a good-quality TTS model can require recording voice actors in a studio-quality acoustic environment and then re-training or fine-tuning the TTS model, which can be laborious and expensive.

This presents a challenge, since the ability to synthesize speech in a rich variety of voices has practical import for many applications.  Designing a TTS model that can generate its own novel voices would be transformative for products such as audiobook readers, speech-based assistants, character voices for games, and video production. It would also present an attractive privacy-preserving alternative to voice cloning systems, which aim to recreate the voice of a real human speaker with a small amount of ground truth audio.

In this paper, we present TacoSpawn, a TTS model that, to the best of our knowledge, is the first system designed to directly generate high-quality speech in voices
not corresponding to a particular human speaker.
We dub this ability  ``speaker generation''.  The main contributions of this work are: 

\begin{itemize}
    \item Proposal and analysis of TacoSpawn, a TTS system with a jointly-trained maximum likelihood estimation model that learns a distribution over speaker embeddings.
    \item Proposal of objective metrics and subjective tests for evaluating speaker generation models. 
    \item Comparison of TacoSpawn against a Bayesian method.    
    \item Experimental results quantifying the strong performance of our approach.
\end{itemize}

\section{Background and related work}
\label{sec:background}

Modern neural text-to-speech (TTS) systems rely on an explicit representation of speaker identity in order to synthesize speech in multiple voices. A good speaker representation uses similar encodings for utterances from the same speaker (even with varying text or prosody), and different encodings for different speakers. When the goal is to synthesize a speaker in the training corpus, this representation may be a simple one-hot encoding of speaker identity \cite{oord2016wavenet}, or embeddings trained jointly with the rest of the model \cite{arik2017deep2}. When the goal is to synthesize an unseen speaker at test time, few-shot (speaker adaptation) or zero-shot (speaker encoder) approaches are often used. In speaker adaptation, all or part of the TTS model is fine-tuned to a small number of audio samples from the unseen speaker \cite{spkadapt_arik2018cloning,spkadapt_spkauxenc_jia2018transfer,spkadapt_moss2020boffin,spkadapt_chen2021adaspeech}. In the speaker encoder approach, embeddings are inferred by an encoder network embedded directly in the TTS model \cite{spkenc_choi2020attentron,spkenc_casanova2021scglowtts}, or using an auxiliary encoder trained on a large amount of audio-only data \cite{spkadapt_spkauxenc_jia2018transfer,spkauxenc_cooper2020zeroshot,spkauxenc_spkenc_Chien2021InvestigatingOI}. The latter may be trained on a speaker-discriminative objective \cite{variani2014dvector,snyder2018xvectors}, or on a voice-conversion task \cite{vc_system_used_by_spkauxenc_spkenc_Chien2021InvestigatingOI}.

Comparatively little research has been devoted to synthesizing speech from truly novel speakers.
Tacotron-2D \cite{spkadapt_spkauxenc_jia2018transfer} conditions a Tacotron
model on an utterance-level \dvector{} at training time, and can be fed uniformly
random points on the unit hypersphere to generate unseen speakers.
However, no attempt is made to ensure that these speakers have a distribution similar to the training speakers, and the use of utterance-level \dvector{}s may limit the extent to which generated speakers can synthesize audio with prosody distinct from that of the reference utterance.
Recent work based on a deep Gaussian process mel spectrogram model 
\cite{mitsui2021gaussianprocess} is similar in spirit to our
approach, sampling from a speaker embedding prior to generate new
speakers. However, postprocessing of the sampled speaker embeddings is required to ensure
that statistics of generated speakers match those of training speakers.
In preliminary experiments, we found that approximately marginalizing over speaker
embeddings using a variational approach typically performed worse
than the TacoSpawn approach described in \sref{sec:model}, especially
when using a standard normal prior as in that work.
It would be interesting to combine the deep Gaussian process mel spectrogram model
with TacoSpawn-style speaker modeling and generation.

\section{Model}
\label{sec:model}

Our approach to speaker generation is conceptually straightforward.
We infer a finite-dimensional \emph{speaker embedding} vector for each speaker in the training
corpus, fit a parametric \emph{speaker embedding prior} distribution to these embeddings, and
sample from this distribution to generate new speakers.
Speakers generated in this way are intended to represent samples from the distribution over human
speakers while not corresponding to any individual human speaker.

In this section we describe this approach in detail.
We review training and synthesis for the \emph{multi-speaker Tacotron} model which underlies
our approach, describe our main speaker generation model \emph{TacoSpawn},
and discuss an alternative variational approach which we do not use in practice.

\subsection{Multi-speaker Tacotron}
\label{sec:tacotron}
The basis of our model is an extension of Tacotron \cite{wang2017tacotron} with trainable
speaker embeddings to support multiple speakers \cite{arik2017deep2}.
The Tacotron model $p_\lam(y | x, s, c)$ autoregressively predicts a mel spectrogram
$y = [y_t]_{t=1}^T$ given a phoneme sequence $x$, a speaker embedding $s \in \reals^D$,
speaker metadata $c$ specifying the speaker's locale and gender, and model parameters $\lam$.
Typically $D \in \{64, 128, 256\}$.
A trainable \emph{speaker embedding table} $S \in \reals^{J \times D}$ specifies the speaker
embedding for each of the $J$ \emph{training speakers} which appear in the training corpus.
The overall \emph{multi-speaker Tacotron} model parameters are $(\lam, S)$.

We train the multi-speaker Tacotron model using maximum likelihood estimation.
Throughout we use $i$ to index the $I$ utterances in the training corpus and $j$ to index the $J$
training speakers.
The training corpus $(Y, X, C)$ consists of target mel spectrograms $Y = [Y_i]_{i=1}^I$,
phoneme sequences $X = [X_i]_{i=1}^I$,
and speaker metadata $C = [C_j]_{j=1}^J$.
We assume it is known which training speaker $j(i)$ produced each utterance $i$.
We learn the speaker embedding table $S = [S_j]_{j=1}^J$ and other model parameters $\lam$ to maximize
the log likelihood
\begin{equation}
    \log p_\lam(Y | X, S, C) = \sum_{i=1}^I \log p_\lam(Y_i | X_i, S_\ji, C_\ji)
\end{equation}
In practice we use the original deterministic Tacotron $\ell_1$ loss when teacher forcing during
training, corresponding to a fixed-variance isotropic Laplace output distribution in the probabilistic
formulation presented here.

A trained multi-speaker Tacotron model can be used to synthesize new phoneme sequences for a training speaker.
Given a phoneme sequence $x$, we generate a mel spectrogram for training speaker $j$ by sampling a mel spectrogram
$y \sim p_\lam(y | x, S_j, C_j)$
with temperature zero, meaning $y_t = \argmax_{y_t} p(y_t | y_{1:t-1}, x, S_j, C_j)$.
We then use a neural vocoder to convert the generated mel spectrogram $y$ to a time-domain waveform
$w = \text{vocode}(y)$ \cite{shen2018natural}.

\subsection{TacoSpawn}
\label{sec:tacospawn}

Our main approach to speaker generation uses the learned speaker embeddings from a multi-speaker Tacotron
model as training data for learning a distribution over speaker embeddings.
In this section we describe this distribution, how to perform maximum likelihood estimation, and how to
use the trained model to generate new speakers.

The \emph{speaker embedding prior} $p_\omega(s | c)$ models the distribution over a speaker embedding
$s \in \reals^D$, given locale and gender metadata $c$ and model parameters $\omega$.
We use a mixture of Gaussians as the parametric form of prior in this work.%
\footnote{
  We use a mixture of Gaussians as a simple and flexible density model
  and make no attempt to interpret the different mixture components as perceptually
  meaningful clusters of speakers.
}
For a prior with $K$ mixture components, we set
\begin{equation}
  p_\omega(s | c) = \sum_{k=1}^K \alpha_{\omega, k}(c)
    \normal\left(s; \mu_{\omega, k}(c), \text{diag}(\sigma^2_{\omega, k}(c))\right)
\end{equation}
where $\omega$ are the parameters of a dense neural net that takes one-hot encodings of the locale and gender
metadata $c$ as input, and produces three outputs: mixture component weights $\alpha_\omega(c) \in \reals^K$ using a softmax activation,
mean vectors $\mu_\omega(c) \in \reals^{K \times D}$, and scale vectors $\sigma_\omega(c) \in \reals^{K \times D}$
using a softplus activation.
Our implementation supports using any subset of $\{\text{locale}, \text{gender}\}$ as the input to the
neural net, including the empty set, in which case we have an \emph{unconditional prior}
$p_\omega(s | c) = p_\omega(s)$.

To train the speaker embedding prior, we use the training speaker embeddings from a trained multi-speaker Tacotron
model as targets.
We learn the parameters $\omega$ to maximize the log likelihood
\begin{equation}
  \log p_\omega(S | C) = \sum_j \log p_\omega(S_j | C_j)
\end{equation}
where $S = [S_j]_{j=1}^J$ is the speaker embedding table learned by a multi-speaker Tacotron model and
$C = [C_j]_{j=1}^J$ is the locale and gender metadata for each training speaker.
We refer to this approach, where a speaker embedding prior is estimated using maximum likelihood on a training speaker
embedding table which was itself estimated using maximum likelihood, as \textbf{TacoSpawn}.

In practice we train $(\lam, S)$ and $\omega$ at the same time, using a stop-gradient operation on
$S$ when optimizing $\log p_\omega(S | C)$ to emulate the separate losses used for
$(\lam, S)$ and $\omega$ above.
Our final training objective is
\begin{equation}
  \label{eq:tmle}
  \begin{aligned}
    L^\text{TacoSpawn}(\lam, \omega, S)
      &= \frac{1}{I} \sum_{i=1}^I \log p_\lam(Y_i | X_i, S_\ji, C_\ji)
  \\
      &\quad + \frac{1}{J} \sum_{j=1}^J \log p_\omega(\text{sg}(S_j) | C_j)
  \end{aligned}
\end{equation}
where $\text{sg}$ is the stop-gradient operation and where the first term is replaced by a minibatch
approximation in practice.%
\footnote{%
    Without the stop-gradient operation, there is a pathological solution to the optimization problem
    with arbitrarily large objective value:
    shrink all speaker embeddings toward a single point and compensate the Tacotron model by adapting
    the scaling of the dense layers which take $s$ as input.
    This makes the prior likelihood arbitrarily large since all the learned speaker embeddings are
    concentrated in a small high-density locale of $\reals^D$.
}
It seems likely that fewer training steps are required to estimate $\omega$ than to estimate $\lam$,
but we found no evidence of overfitting of $\omega$ when using a larger number of training steps
appropriate for learning $\lam$.%
\footnote{%
  We tested for overfitting by performing experimental runs where only half of the training
  speakers were included in the data used to train the speaker embedding prior, with the
  other half used as an eval set.
  For both halves the speaker embeddings were learned to maximize the likelihood under
  the multi-speaker Tacotron model as usual.
  The speaker embedding prior log prob on the eval half of speakers closely matched the
  log prob on the training half of speakers throughout training, suggesting no overfitting.
}

A trained speaker embedding prior can be used to generate new speakers.
Given desired locale and gender metadata $c$ for a new speaker, we generate a speaker embedding by sampling
$s \sim p_\omega(s | c)$ with temperature one.
We can synthesize speech from the generated speaker using the approach described in \sref{sec:tacotron},
generating a mel spectrogram by sampling $y \sim p_\lam(y | x, s, c)$ with temperature zero.

In the TacoSpawn approach, the prior does not affect estimation of the training speaker embeddings.  This means there is no incentive for the embeddings to have a particular distribution, which informs our choice of a fairly generic and flexible form of prior. Despite its simplicity, we find that a mixture of Gaussians results in high-quality speaker generation.

\subsection{\tacovb{}}
\label{sec:vb}
An alternative approach to speaker generation is to treat the speaker embedding prior and Tacotron model
as part of one larger model, treating the speaker embedding for each speaker as a latent variable to be
inferred as part of training.
This corresponds to learning the speaker embedding table $S$ using a Bayesian approach instead of using
maximum likelihood estimation.
We specifically consider a variational Bayesian \cite{attias1999variational} approach with
non-amortized stochastic variational inference \cite{hoffman2013stochastic}
and refer to this formulation as \textbf{\tacovb{}}.

We train the speaker embedding prior $p_\omega(s | c)$ and Tacotron model $p_\lam(y | x, s, c)$
together to approximately maximize the likelihood of the training corpus $(Y, X, C)$.
For any \emph{variational posterior} distribution $q_\nu(S)$ over the speaker embedding table $S$,
the log marginal likelihood has lower bound
\begin{align}
    &\log p_{(\lam, \omega)}(Y | X, C)
\\
    &= \log \int p_\lam(Y | X, S, C) p_\omega(S | C) \,dS
\\
    &= \log \int \frac{p_\lam(Y | X, S, C) p_\omega(S | C)}{q_\nu(S)} q_\nu(S) \,dS
\\
  \label{eq:elbo1}
    &\begin{aligned}
        &\geq \int q_\nu(S) \log p_\lam(Y | X, S, C) \,dS
      \\
        &\quad + \int q_\nu(S) \left[ \log p_\omega(S | C) - \log q_\nu(S) \right] \,dS
    \end{aligned}
\end{align}
with equality if and only if $q_\nu(S) = p_{(\lam, \omega)}(S | Y, X, C)$ for all $S$.
Maximizing \eqref{eq:elbo1} with respect to $(\lam, \omega, \nu)$ therefore performs approximate
maximum likelihood estimation of $(\lam, \omega)$, where learning $\nu$ helps make the lower bound
as tight as possible.
Since the true \emph{generative posterior} over speaker embeddings factorizes across training speakers,
that is
\begin{equation}
  p_{(\lam, \omega)}(S | Y, X, C) = \prod_{j=1}^J p_{(\lam, \omega)}(S_j | Y, X, C_j)
\end{equation}
we can assume without loss of generality that $q_\nu(S) = \prod_j q_{\nu_j}(S_j)$
where $\nu_j$ are the parameters of the distribution for training speaker $j$.
The lower bound \eqref{eq:elbo1} becomes
\begin{equation}
  \label{eq:vb-strict}
  \begin{aligned}
      &\sum_{i=1}^I \int q_{\nu_\ji}(S_\ji) \log p_\lam(Y_i | X_i, S_\ji, C_\ji) \,dS_\ji
    \\
      &\quad + \sum_{j=1}^J
        \int q_{\nu_j}(S_j) \left[ \log p_\omega(S_j | C_j) - \log q_{\nu_j}(S_j) \right] \,dS_j
  \end{aligned}
\end{equation}
We use a diagonal Gaussian as the parametric form of variational posterior in this work, that is
\begin{equation}
  q_{\nu_j}(s) = \normal\left(s; \mu_j, \text{diag}(\sigma^2_j)\right)
\end{equation}
where $\mu, \sigma \in \reals^{J \times D}$ and $\nu_j = (\mu_j, \sigma_j)$.
As in \sref{sec:tacospawn} we use a softplus activation to ensure that each component of the scale
vector $\sigma_j$ is positive.

Following the beta-VAE \cite{higgins2016beta}, we introduce an additional parameter $\beta$
to control how much information speaker embeddings contain about the way a speaker speaks.
Our final training objective is
\begin{equation}
  \label{eq:vb}
  \begin{aligned}
    &L^\text{\tacovb{}}(\lam, \omega, \nu)
  \\
      &= \frac{1}{I} \sum_{i=1}^I \int q_{\nu_\ji}(S_\ji) \log p_\lam(Y_i | X_i, S_\ji, C_\ji) \,dS_\ji
  \\
      &\quad + \beta \frac{1}{J} \sum_{j=1}^J
        \int q_{\nu_j}(S_j) \left[ \log p_\omega(S_j | C_j) - \log q_{\nu_j}(S_j) \right] \,dS_j
  \end{aligned}
\end{equation}
where $\beta = J / I$ for a strict Bayesian formulation as in \eqref{eq:vb-strict}.
In practice the first term is replaced by a minibatch approximation and each expectation
over $S_\ji$ and $S_j$ is approximated using a single reparameterized sample.
We refer to the negative of the term which $\beta$ multiplies as the \emph{KL term}.
The KL term is an upper bound on the \emph{capacity} of the speaker embedding \cite{higgins2016beta},
which here means the representational mutual information between a speaker's collection of $(x, y)$ pairs
and their speaker embedding $s$ (conditional on any metadata that the prior is conditioned on).
During training, we automatically tune $\beta$ to target a particular value of the KL term
\cite{rezende2018taming,battenberg2019capacitron}, thus bounding the capacity and so controlling
how much information speaker embeddings contain about the text and audio produced by a speaker.

There is an important implementation detail related to how we average in \eqref{eq:vb}.
The training objective consists of one term which is an average over utterances, and one term
which is an average over speakers.
It is crucial for a proper mathematical formulation that the second term is an average
over all speakers in the training corpus rather than the speakers in the current minibatch.
In preliminary experiments we found the more principled approach worked better in practice.

A Bayesian approach has a number of potential advantages:
it naturally integrates learning the Tacotron model parameters $\lam$ and the prior parameters
$\omega$ in a coherent way;
it prevents overfitting of the speaker embedding for speakers with few utterances;
and it naturally encourages the learned embeddings to be easily modelable under the parametric prior,
in contrast to the TacoSpawn approach where the prior has no effect on the learned embeddings.
Nevertheless we find \tacovb{} to perform no better than TacoSpawn empirically
(\sref{sec:prelim-expt}).

\subsection{TacoSpawn as a limiting case of \tacovb{}}
\label{sec:beta-zero}
We can view TacoSpawn as the high-capacity limit of the more principled \tacovb{} approach.
As $\beta \to 0$ in \eqref{eq:vb}, the training dynamics of \tacovb{} become essentially the
same as those of TacoSpawn.
In this section we establish this result, first considering the objective used to learn
$\lam$ and $S$ or $\nu$ and then considering the objective used to learn $\omega$.

The TacoSpawn objective for learning $\lam$ and $S$ is
\begin{equation}
  \frac{1}{I} \log p_\lam(Y | X, S, C)
\end{equation}
As $\beta \to 0$, the \tacovb{} objective for learning $\lam$ and $\nu$ tends to
\begin{equation}
  \label{eq:vb-beta-zero}
  \frac{1}{I} \int q_\nu(S) \log p_\lam(Y | X, S, C) \,dS
\end{equation}
Since \eqref{eq:vb-beta-zero} is the expected value of some function of $S$ with respect to $q_\nu(S)$,
the optimal $q_\nu$ puts all its mass on a single value, namely $\argmax_S \log p_\lam(Y | X, S, C)$,
so it is not a restriction to assume that
$q_\nu$ is deterministic, in which case the \tacovb{} objective for $\theta$ and $\nu$ reduces to the
TacoSpawn objective for $\theta$ and $S$.

The TacoSpawn objective for learning $\omega$ is
\begin{equation}
  \frac{1}{J} \log p_\omega(S | C)
\end{equation}
The \tacovb{} objective for learning $\omega$ is
\begin{equation}
  \beta \frac{1}{J} \int q_\nu(S) \log p_\omega(S | C) \,dS
\end{equation}
If we use an optimizer such as Adam \cite{kingma2015adam} which is approximately invariant to loss scaling,
or if we use stochastic gradient descent and scale the learning rate on $\omega$ by $1 / \beta$ (likely a good
idea in any case from an optimization standpoint), then we may effectively drop the $\beta$ multiplier to obtain
\begin{equation}
  \frac{1}{J} \int q_\nu(S) \log p_\omega(S | C) \,dS
\end{equation}
which reduces to the TacoSpawn objective for $\omega$ when $q_\nu$ is deterministic.

\section{Evaluating speaker generation}
\label{sec:methods_speaker_distance}
To evaluate a speaker generation system, we'd like to measure both how realistic-sounding and diverse generated speakers are. These attributes can be difficult to quantify explicitly, however, and evaluating some aspects of diversity is tricky even for human listeners.\footnote{
  For example, if a human listened to $1000$ speaker samples, we wouldn't expect him/her to realize (without a side-by-side comparison) that any two speakers sounded disproportionately similar to each other.
}
In this section we address these challenges, and propose intuitive objective measures of system performance.
Our general methodology is to look for deficiencies in the speaker generation process by comparing
statistics of the probability distributions of training and generated speakers' audio.
For a well-trained speaker generation system these two distributions would be identical, and so any
audio-based statistic whose value differs between training and generated speakers indicates a flaw
in speaker generation.
We base our statistics on
\emph{\dvector{}s} \cite{variani2014dvector} to encourage them to be sensitive
to speaker identity \cite{wang2017dvector_ivector_comparison}.

\subsection{Speaker-level \dvector{}s}
\label{subsec:methods_speaker_realizations}
We first describe how we compute speaker-level \dvector{}s based on training and generated
speakers' audio.
We assume we have access to an evaluation corpus $(W^\text{eval}, X^\text{eval})$ containing
ground truth audio waveforms $W^\text{eval} = [W^\text{eval}_i]_{i=1}^{I^\text{eval}}$ and
phoneme sequences $X^\text{eval} = [X^\text{eval}_i]_{i=1}^{I^\text{eval}}$ for $I^\text{eval}$
additional utterances from the training speakers.
Let $I^\text{eval}(j) = \{i \in \{1, \ldots, I^\text{eval}\} : j(i) = j\}$ be the set of
evaluation utterances produced by speaker $j$.
Let $\overline{v}$ be a \dvector{} model which takes a speech waveform $w$ as input and returns a unit-norm
vector $\overline{v}(w) \in \reals^{256}$ intended to represent speaker-specific information
present in the waveform.
We compute speaker-level \dvector{}s on three different versions of the evaluation corpus:
\begin{itemize}
\item[(t)]
    \emph{Ground truth training speaker audio}.
    We compute a \dvector{} $V^\text{t}_j$ for each training speaker $j$
    by averaging utterance-level
    \dvector{}s computed on their ground truth audio:
    \begin{align*}
      V^\text{t}_j &= \frac{1}{|I^\text{eval}(j)|} \sum_{i \in I^\text{eval}(j)}
        \overline{v}(W^\text{eval}_i), & j &= 1, \ldots, J
    \end{align*}
\item[(s)]
    \emph{Synthesized training speaker audio}.
    We synthesize the evaluation corpus using the inferred training speaker embeddings and
    similarly compute a \dvector{} $V^\text{s}_j$ for each
    training speaker $j$:
    \begin{align*}
      S_j &\sim q_{\nu_j}(s), & j &= 1, \ldots, J
    \\
      Y^\text{s}_i &\sim p_\lam(y | X^\text{eval}_i, S_\ji, C_\ji), & i &= 1, \ldots, I^\text{eval}
    \\
      W^\text{s}_i &= \text{vocode}(Y^\text{s}_i), & i &= 1, \ldots, I^\text{eval}
    \\
      V^\text{s}_j &= \frac{1}{|I^\text{eval}(j)|} \sum_{i \in I^\text{eval}(j)} \overline{v}(W^\text{s}_i), & j &= 1, \ldots, J
    \end{align*}
    where the mel spectrogram $Y^\text{s}_i$ is sampled with temperature zero (\sref{sec:tacotron}).
    For TacoSpawn models, $S$ is deterministically set to the learned speaker embedding table.
\item[(g)]
    \emph{Synthesized generated speaker audio}.
    We synthesize the evaluation corpus using generated speaker embeddings, one generated
    speaker per training speaker, and compute a \dvector{} $V^\text{g}_j$ for each generated
    speaker $j$:
    \begin{align*}
      S^\text{g}_j &\sim p_\omega(s | C_j), & j &= 1, \ldots, J
    \\
      Y^\text{g}_i &\sim p_\lam(y | X^\text{eval}_i, S^\text{g}_\ji, C_\ji), & i &= 1, \ldots, I^\text{eval}
    \\
      W^\text{g}_i &= \text{vocode}(Y^\text{g}_i), & i &= 1, \ldots, I^\text{eval}
    \\
      V^\text{g}_j &= \frac{1}{|I^\text{eval}(j)|} \sum_{i \in I^\text{eval}(j)} \overline{v}(W^\text{g}_i), & j &= 1, \ldots, J
    \end{align*}
\end{itemize}
Using the same number of speakers and speaker metadata for training and generated speakers helps
ensure that a difference between statistics computed on training and generated speakers
indicates a flaw in speaker generation.

\subsection{Speaker distance metrics}
\label{subsec:methods_speaker_distance_metrics}
In this section we propose intuitive objective measures of both speaker generation performance and training speaker fidelity. We use the naming convention $x$2$y$ to denote the distance from $x$ to $y$, where $x$ and $y$ are one of
t (ground truth training speaker audio),
s (synthesized training speaker audio) or
g (synthesized generated speaker audio).

To evaluate speaker generation performance, writing
$d(v_1, v_2) = 1 - \frac{v_1}{\lVert v_1 \rVert} \cdot \frac{v_2}{\lVert v_2 \rVert}$ for the cosine distance
\cite{dehak2011front}, we compute:
\begin{itemize}
\item 
  \textbf{s2s}: How close is a typical training speaker to other nearby training speakers, when
  both are synthesized?
  \begin{equation}
    \median_j \min_{k \neq j} d(V^\text{s}_j, V^\text{s}_k)
  \end{equation}
\item 
  \textbf{g2s}: How close is a typical generated speaker to nearby training speakers, where both are synthesized?
  (We exclude $k = j$ for uniformity).
  \begin{equation}
    \median_j \min_{k \neq j} d(V^\text{g}_j, V^\text{s}_k)
  \end{equation}
\item 
  \textbf{g2g}: How close is a typical generated speaker to other nearby generated speakers?
  \begin{equation}
    \median_j \min_{k \neq j} d(V^\text{g}_j, V^\text{g}_k)
  \end{equation}
\end{itemize}
We can compare s2s, g2s, and g2g to evaluate speaker generation performance.
For an ideal system, the statistics of training and
generated speakers' synthesized audio will be identical, and so s2s, g2s and g2g
will be equal.
Comparing g2s to s2s helps detect whether generated speakers sound disproportionately dissimilar
to training speakers ($\text{g2s} > \text{s2s}$), i.e.\ whether generated speakers' audio lies
off the manifold of training speakers' audio. This provides a measure of how \emph{realistic}
or \emph{natural} the generated speakers are.
Comparing g2s to s2s also helps to detect whether generated speakers are disproportionately
similar to the training speakers ($\text{g2s} < \text{s2s}$), indicating \emph{overfitting}.
Comparing g2g to s2s helps detect whether generated speakers sound 
disproportionately similar to each other ($\text{g2g} < \text{s2s}$), and thus provides
a measure of the \emph{diversity} of generated speakers.
For example, if all generated speakers individually sounded like a plausible new speaker, but basically
all sounded the same, then g2s would be close to s2s (good) but g2g would be very small (bad).

Speaker generation performance cannot be entirely disentangled from the fidelity with which the
model captures training speaker characteristics.
We therefore also evaluate \emph{speaker fidelity} by computing:
\begin{itemize}
\item 
  \textbf{s2t-same}: How similar is synthesized audio from a typical training speaker to ground truth audio from that speaker?
  \begin{equation}
    \median_j d(V^\text{s}_j, V^\text{t}_j)
  \end{equation}
\item 
  \textbf{s2t}: How similar is synthesized audio from a typical training speaker
  to ground truth audio from other nearby training speakers?
  \begin{equation}
    \median_j \min_{k \neq j} d(V^\text{s}_j, V^\text{t}_k)
  \end{equation}
\end{itemize}
The smaller the value of s2t-same, the closer synthetic speech from a training speaker is to their natural speech.%
\footnote{%
    Note that \dvector{}s vary substantially across utterances from the same
    speaker for both natural and synthesized speech.
    For example, preliminary Griffin-Lim experiments showed an average distance of 0.15 
    between the speaker-level \dvector{}s computed on audio synthesized from the same speaker on two 
    different collections of text.
    We therefore do not expect s2t-same to be zero even for a perfect multi-speaker
    speech synthesis system.
}
We use s2t as a dataset-specific reference point against which to compare s2t-same.
A potential system weakness detected by s2t-same but not by (s2s, g2s, g2g) is where
the speaker identity is muddy or indistinct in synthesized audio from both
training and generated speakers.

For the system taken as a whole, we look for g2s and g2g equal to s2s to indicate effective speaker generation and s2t-same as small as possible relative to s2t to indicate strong training speaker fidelity.

\section{Experimental setup}
\label{subsec:datasets}
In the sections that follow, we show results using both public and proprietary multi-speaker English datasets, all sampled at 24~kHz.

\begin{itemize}
    \item \textbf{libriclean} (public): All ``clean'' subsets of the LibriTTS corpus, combined into one (US English, 1230 speakers, \textapprox{240.5} hours, mixed-gender).
    \item \textbf{enus1100} (proprietary): A 1100-speaker US English dataset of mixed-gender voices speaking for 30 minutes each, for a total of \textapprox{246,000} utterances (\textapprox{500} hours). We evaluate on $2\%$ of these utterances. %
    \item \textbf{en1468} (proprietary): A 1468-speaker English dataset that augments enus1100 with voices in four locales (US, British, Australian, and Indian English). These include audiobook, voice assistant, and news-reading English speakers, many of which have been used in previous research \cite{oord2016wavenet,wang2017tacotron,shen2018natural}. These total \textapprox{876,000} training utterances (\textapprox{717} hours) from 1468 speakers. We evaluate on $1\%$ of these utterances. 
\end{itemize}
We train our models on input phoneme sequences produced by a text normalization front-end and lexicon, since our focus is speaker representation rather than the model's ability to learn pronunciations from graphemes.  
We optimize our models using Adam \cite{kingma2015adam} for \textapprox{300k} steps on 32 Google TPUv3 cores, using batch size 256.
The \dvector{} model used for speaker distance metrics, t-SNE plots, and the \dvector{} prior in \sref{sec:dvector-prior}
has 256-dimensional output and is trained on a separate corpus with a speaker-discriminative loss.

\section{Preliminary experiments}
\label{sec:prelim-expt}

We first report results of preliminary experiments that informed our choice of approach. All models in this section were trained on the enus1100 dataset described in \sref{subsec:datasets}, and used Griffin-Lim vocoding to allow fast iteration.

\begin{table}[t]
    \centering
    \begin{tabular}{l c c c c c c c}
        \toprule
        \multirow{2}{*}{approach} & \multirow{2}{*}{\multilinecell{c}{KL\\term}} & \multicolumn{2}{c}{speaker fidelity} & \multicolumn{3}{c}{speaker generation} \\
        \cmidrule(lr){3-4} \cmidrule(lr){5-7}
        & & s2t-same & s2t & s2s & g2s & g2g \\
        \midrule
        TacoSpawn & & 0.22 & 0.34 & 0.20 & 0.20 & 0.20 \\
        TS-VB & 898 & 0.23 & 0.34 & 0.20 & 0.20 & 0.19 \\
        TS-VB & 400 & 0.23 & 0.34 & 0.19 & 0.19 & 0.19 \\
        TS-VB & 100 & 0.25 & 0.34 & 0.18 & 0.17 & 0.16 \\
        TS-VB & 40 & 0.29 & 0.33 & 0.15 & 0.14 & 0.14 \\
        TS-VB & 20 & 0.33 & 0.32 & 0.13 & 0.13 & 0.12 \\
        \bottomrule
    \end{tabular}
    \caption{%
        Evaluation of the variational Bayes (TS-VB) approach.
        A KL term value of $898$ corresponds to $\beta = 10^{-5}$.
    }
    \label{table:tmle-vs-vb}
\end{table}

\subsection{\tacovb{}}
We first compared the \tacovb{} approach described in \sref{sec:vb} to the TacoSpawn approach described
in \sref{sec:tacospawn}.
We used a $128$-dimensional speaker embedding and a learned unconditional mixture of $10$ Gaussians prior
in all cases.
\tabref{table:tmle-vs-vb} shows objective results for these experiments.
As expected from \sref{sec:beta-zero}, \tacovb{} performed similarly to TacoSpawn for large KL term (small $\beta$).
However as we decreased the KL term (increase $\beta$), speaker fidelity (s2t-same) degraded.
This led us to favor the simpler TacoSpawn approach.

\begin{table}[t]
    \centering
    \begin{tabular}{l c c c c c c c}
        \toprule
        \multirow{2}{*}{approach} & \multirow{2}{*}{dim} & \multicolumn{2}{c}{speaker fidelity} & \multicolumn{3}{c}{speaker generation} \\
        \cmidrule(lr){3-4} \cmidrule(lr){5-7}
        & & s2t-same & s2t & s2s & g2s & g2g \\
        \midrule
        TacoSpawn & 128 & %
            0.23 & 0.34 & 0.20 & 0.20 & 0.20 \\
        TacoSpawn & 256 & %
            0.22 & 0.34 & 0.20 & 0.23 & 0.18 \\
        \dvector{} & 256 & %
            0.23 & 0.34 & 0.20 & 0.35 & 0.27 \\
        \bottomrule
    \end{tabular}
    \caption{%
        Preliminary experiments investigating using speaker-level \dvector{}s as a speaker embedding.
    }
    \label{table:tmle-vs-dvec}
\end{table}

\subsection{Learned prior over \dvector{}s}
\label{sec:dvector-prior}
Due to the success of \dvector{}s for voice cloning
 \cite{spkadapt_spkauxenc_jia2018transfer}, we also investigated 
using \dvector{}s as speaker embeddings.
For each training speaker, we computed a speaker-level \dvector{} by averaging their
utterance-level \dvector{}s computed on training corpus ground truth audio.
These speaker-level \dvector{}s were then used as the training speaker embeddings when training
the Tacotron model and the speaker embedding prior, which was an
unconditional mixture of $10$ Gaussians.
Objective results are shown in \tabref{table:tmle-vs-dvec}.
We can see that \dvector{}s capture the training speakers reasonably well (s2t-same), though still not as well as learned vectors (TacoSpawn) of the same dimension.
However \dvector{}s appear to be much less amenable to modeling with a simple parametric prior (g2s and g2g much too large).
We therefore use learned vector embeddings.

\section{TacoSpawn Experiments}
\label{sec:results}

We now show experimental results for our proposed TacoSpawn model. We trained these on the libriclean, en1468, and enus1100 datasets, and used an adversarially-trained non-causal WaveNet vocoder (full parameters in Appendix \ref{appendix:table:tacotron_hparams}).

\subsection{Objective evaluation results}

\subsubsection{Speaker distance metrics}
\label{subsec:speaker_distance}

To evaluate how well the distribution of generated speakers matches that of speakers in the training corpus, we calculated the speaker distance metrics proposed in \sref{subsec:methods_speaker_distance_metrics}.  \tabref{table:speaker-distance-metrics} shows results for both public and proprietary dataset models.
We can see that s2t-same is lower than s2t, providing a useful sanity check that the model
successfully captures characteristics of speaker identity for the training speakers.
We also see that 
g2s and g2g almost perfectly match s2s, indicating that generated speakers are statistically as diverse and realistic as training speakers.

As an aside, we note that s2t-same is larger than s2s for many models. While not ideal, this reflects the fact that waveforms output by neural TTS models are still distinguishable from ground truth audio, albeit not necessarily by humans. Specifically, the \dvector{} model we use to calculate speaker distance metrics may be sensitive to acoustic characteristics such as reverb, channel noise, or speech synthesis artifacts. \cite{spkadapt_spkauxenc_jia2018transfer, jia2021translatotron2}
Since the focus of this work is achieving speaker generation rather than solving speech synthesis, this does not hinder our analysis.

\begin{table}[t]
    \centering
    \begin{tabular}{l c c c c c}
        \toprule
        \multirow{2}{*}{dataset} & \multicolumn{2}{c}{speaker fidelity} & \multicolumn{3}{c}{speaker generation} \\
        \cmidrule(lr){2-3} \cmidrule(lr){4-6}
        & s2t-same & s2t & s2s & g2s & g2g \\
        \midrule
        libriclean  & 0.42  & 0.51 & 0.41 & 0.41 & 0.40 \\ %
        en1468      & 0.27  & 0.33 & 0.17 & 0.18 & 0.17 \\ %
        enus1100    & 0.14	& 0.30 & 0.24 &	0.24 & 0.24 \\
        \bottomrule
    \end{tabular}
    \caption{%
Speaker distance metrics for TacoSpawn models.
    }
    \label{table:speaker-distance-metrics}    
\end{table}

\subsubsection{Visualizing speaker distance}
\label{subsubsec:tsne_speaker_distance}

\begin{figure}
    \centering
    \includegraphics[width=0.5\textwidth]{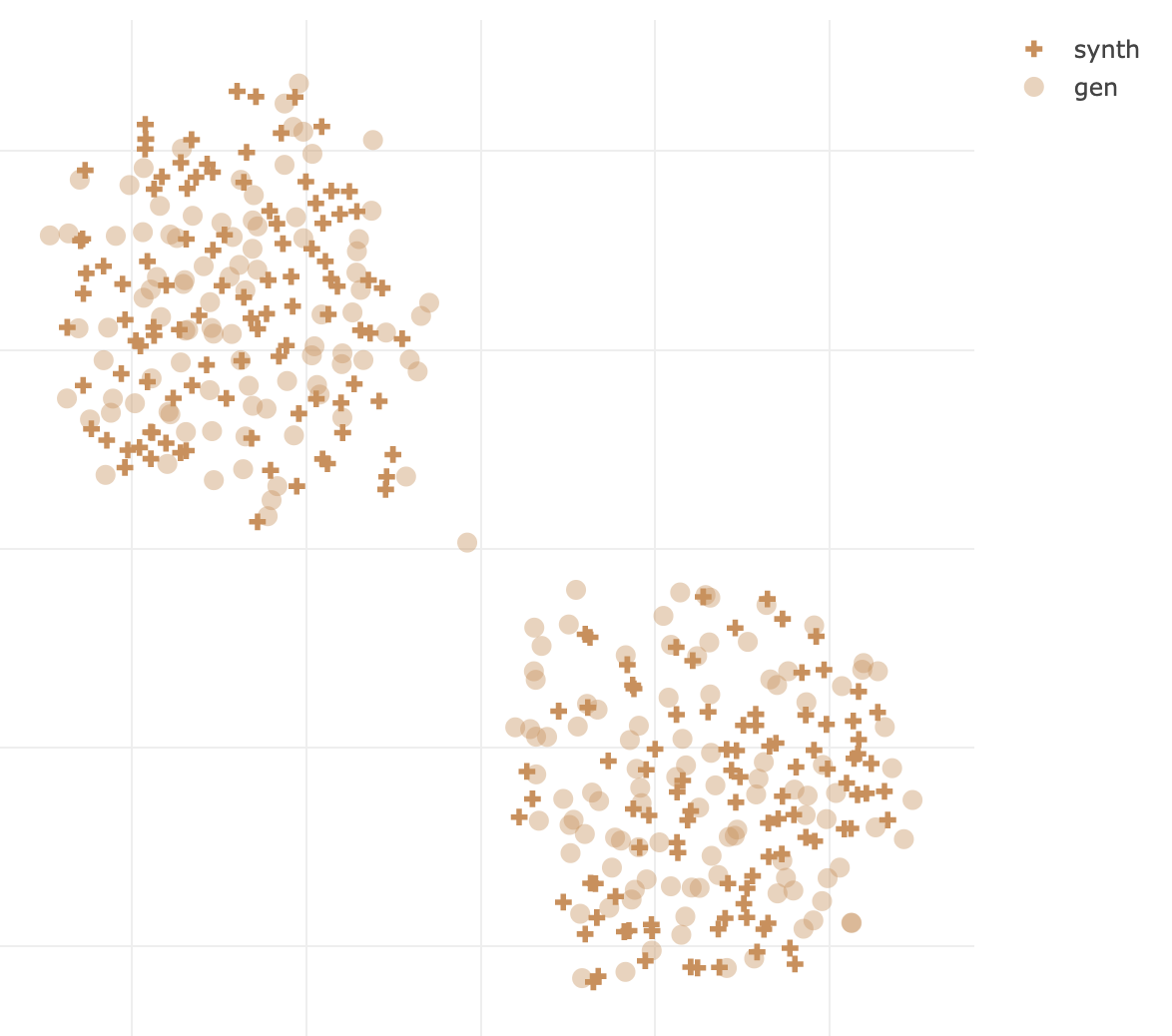}
    \caption{t-SNE plot of speaker \dvector{}s for 100 synthesized and generated speakers from our libriclean TacoSpawn model.}
    \label{fig:ex184c3_enus_100_per_lrg_quickbrownfox_dvector_tsne}
\end{figure}

Plotting speaker \dvector{}s helps visualize the speaker distance relationships discussed in \sref{subsec:speaker_distance}. Figure~\ref{fig:ex184c3_enus_100_per_lrg_quickbrownfox_dvector_tsne} shows t-SNE plots of \dvector{}s extracted from audio synthesized by our libriclean model for both training and generated speakers. As expected, the \dvector{}s fall into two gender clusters.
We can see that the training and generated speaker distributions are nicely overlapping, and it does not appear that generated speakers clump disproportionately close to training speakers.
See our audio demo page for an interactive t-SNE plot of \dvector{}s extracted from multi-locale audio synthesized by our en1468 model.

\subsubsection{Fundamental frequency analysis}
\label{subsec:pitch}

\begin{figure}
    \centering
    \includegraphics[width=0.5\textwidth]{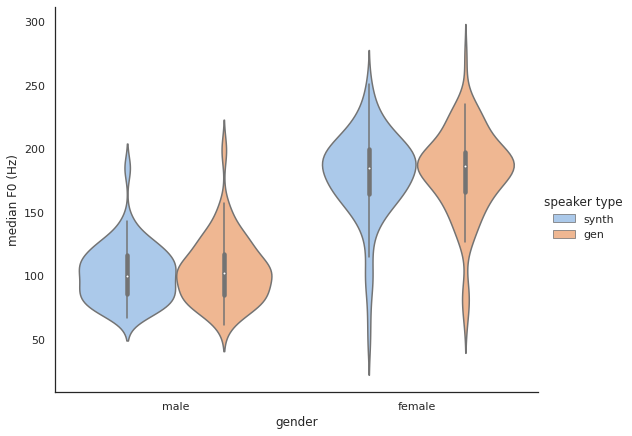}
    \caption{Median fundamental frequency of 200 training (synth) and 200 generated (gen) speakers for audio synthesized using the libriclean TacoSpawn model, split by gender.}
    \label{fig:ex184c4_libriclean_mog_mgv3_ftlibrimops_enus_100_per_lrgt_libricorpus_F0_violinplot_v2}
\end{figure}

To further examine the perceptual properties captured by speaker distance metrics, we also compared the fundamental frequency ranges of synthesized audio from training and generated speakers,
following Mitsui et al.\ \cite{mitsui2021gaussianprocess}.
We used the Yin \cite{decheveigne2002yin} extraction algorithm with a frame shift of 12.5~ms to compute median fundamental frequency of utterances spoken by 200 generated and 200 training speakers (100 of each gender).  The results, plotted in Figure~\ref{fig:ex184c4_libriclean_mog_mgv3_ftlibrimops_enus_100_per_lrgt_libricorpus_F0_violinplot_v2}, show that the fundamental frequency range of training and generated speakers are equally diverse, clearly cluster into male and female, and are distributionally similar.

\subsection{Subjective evaluation results}

\subsubsection{Subjective speaker similarity}
\label{subsec:human_eval_speaker_similarity_correlation} 

To examine the degree to which the \dvector{}--based speaker similarity metrics proposed in \sref{subsec:methods_speaker_distance_metrics} capture speaker identity characteristics, we asked a pool of human reviewers to determine (boolean yes or no) whether 1294 utterance pairs were uttered by the same speaker. Each utterance was 3-5 seconds long, and either ground truth audio (t), synthesized training speaker audio (s), or synthesized generated speaker audio (g). Transcripts used for evaluation were unseen during training and typically different within a pair. The TacoSpawn model was trained on the enus1100 dataset
and used a WaveRNN \cite{vocoder_wavernn} vocoder.  We included \textapprox{220} of each of the six possible utterance pair types (g2g, g2s, s2s, g2t, s2t, t2t), and drew pairs from the entire range of \dvector{} scores. 

\begin{figure}
    \includegraphics[width=0.5\textwidth]{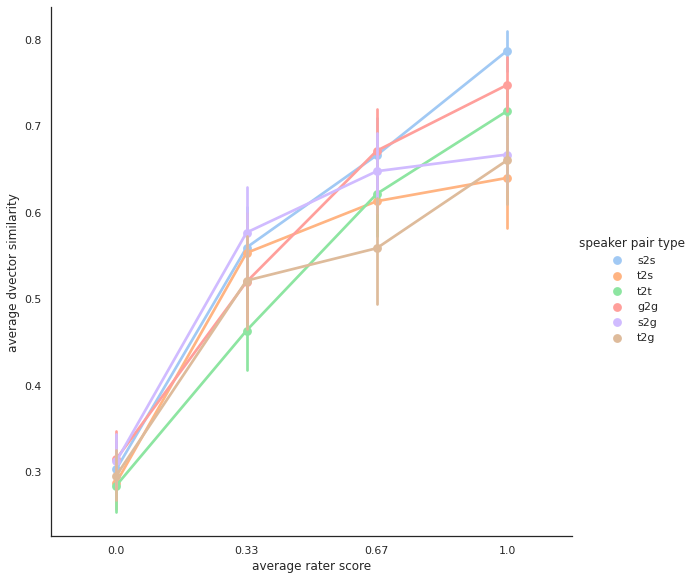}
    \caption{Average correlation between \dvector{} cosine similarity ($y$ axis) and human ``same speaker'' ratings ($x$ axis) of pairs of US English utterances, broken down by speaker pair type. The $x$ axis ticks reflect the average of 3 boolean ratings per utterance pair.}
    \label{fig:human_eval_speaker_similarity_correlation}
\end{figure}

Figure~\ref{fig:human_eval_speaker_similarity_correlation}
shows the correlation between \dvector{} cosine similarity and the average human ratings, and indicates that the \dvector{}--based metric captures some perceptual notion of speaker identity.

\subsubsection{Speech naturalness}
\label{subsec:mos} 

We measured the naturalness of speaker generation models in three English locales using subjective listening tests. For each model and locale, we randomly selected a fixed number of gender-balanced training and generated speakers, and synthesized one utterance per speaker.
A pool of human reviewers rated the naturalness of these utterances on a scale from 1 (bad) to 5 (excellent) in increments of 0.5 \cite[Figure A.2]{battenberg2019capacitron}. 
Each utterance received three independent ratings, and reviewers rated no more than \textapprox{25} utterances each. Results are shown in Table~\ref{table:mos_naturalness_enxx}.  Mean opinion scores (MOS) for training and generated speakers are similar, indicating that TacoSpawn voices achieve quality comparable to speakers in the training corpus. We encourage readers to visit our demo page to listen to audio samples.

\begin{table}[t]

    \centering
    \begin{tabular}[t]{l c c c c c} %
        \toprule
        data     & locale & %
        spkrs
        & \multilinecell{c}{training\\speakers} & \multilinecell{c}{generated\\speakers} \\ 
        \midrule
        libriclean  & us    & 200   & \mosv{3.37}{0.14}       %
                                    & \mosv{3.54}{0.14}\\  %
        en1468        & au    & 164   & \mosv{3.30}{0.15}     %
                                        & \mosv{3.03}{0.14}\\   %
        en1468        & us    & 300   & \mosv{3.68}{0.11}        %
                                    & \mosv{3.62}{0.11} \\     %
        en1468        & gb    & 212   & \mosv{3.69}{0.12}        %
                                    & \mosv{3.51}{0.13} \\     %
        \bottomrule
    \end{tabular}
    \caption{%
        Speech naturalness mean opinion score (MOS) of utterances synthesized by systems trained on the libriclean and en1468 datasets, broken down by %
        locale. Scores are shown for both training and generated speakers with 95\% confidence intervals.
    }
    \label{table:mos_naturalness_enxx}    
\end{table}

\section{Conclusions}
\label{sec:conclusion} 

This work explored speaker generation, the ability of a model to generate speech in novel, human-sounding voices unseen in any training corpus. We proposed TacoSpawn, a method for learning a distribution over a speaker embedding space that can be efficiently co-trained with an existing TTS model.  We presented a set of statistics to evaluate speaker generation performance, developed intuition for these metrics, discussed experimental results that informed our choice of approach, and showed experimental results on both public and proprietary datasets. To the best of our knowledge, TacoSpawn is the first self-contained TTS model that can directly generate high-quality speech from novel speakers. 

We hope the explicit formulation of the speaker generation task inspires further research.
Encouragingly, there are many directions to explore, since the models described in this paper extend only a basic multi-speaker Tacotron. 
Combining speaker generation with learned prosody representations \cite{skerry2018towards, wang2018gst, stanton2018tpgst, battenberg2019capacitron} may improve the expressiveness of generated speakers. Experimenting with audio-only training data may allow learning speaker representations from a much larger corpus of public-domain data. It would also be interesting to combine other speech synthesis architectures with TacoSpawn-style speaker modeling and generation.
We expect that any of these approaches will be exciting areas for future work.

\section{Acknowledgments}

The authors would like to thank Rif A. Saurus, Ye Jia, Rob Clark, Chun-an Chang, Yu Zhang, and Heiga Zen and for their helpful feedback and discussion.

\bibliographystyle{IEEEbib}
\bibliography{refs}

\clearpage{}%

\newcommand{\rowheader}[2]{\multirow{#1}{15mm}{#2}}

\newpage
\appendix
\newpage

\begin{center}
    \large{\bf{Appendix}}
\end{center}

\setcounter{figure}{0}    
\setcounter{table}{0}    

\section{Neural Network Architecture}\label{appendix:architecture}

In this section we provide a detailed description of the speech synthesis system to which we added TacoSpawn speaker generation. Some of this reference material is copied from our previous publications. \cite{habib2020semisupervisedtacotron}

\paragraph{Sequence-to-sequence model}
Our mel-spectrogram prediction network is based on Tacotron \cite{wang2017tacotron} with minor modifications from follow-up work \cite{skerry2018towards}. Input to the model consists of sequences of phonemes produced by a text normalization pipeline rather than character inputs. The CBHG text encoder from \cite{wang2017tacotron} is used to convert the input phonemes into a sequence of text embeddings. The phoneme inputs are converted to learned 256-dimensional embeddings and passed through a pre-net composed of two fully connected ReLU layers (with 256 and 128 units, respectively), with dropout of 0.5 applied to the output of each layer, before being fed to the encoder. The learned $128$-dimensional speaker embedding is broadcast-concatenated to the output of the text encoder.  

\begin{table}[h!]
    \centering
    \begin{tabular}{p{18mm} l}
    \toprule
    Module & hyperparameters    \\
    \midrule
     Input & Text-normalized phonemes \\
     Phoneme embedding  & 256-D \\
     Pre-net & FC-256-Relu-Dropout(0.5) \\
          & $\rightarrow$ FC-128-Relu-Dropout(0.5) \\
     \cellv{CBHG}{text encoder} &  Conv1D bank: K=16, conv-k-128-Relu  \\                         &  $\rightarrow$Max pooling with stride=1  width=2  \\
          & $\rightarrow$ Conv1D projections: conv-3-128-Relu  \\
          & $\rightarrow$ conv-3-128-Linear\\
          & $\rightarrow$ Highway net: 4 layers of FC-128-Relu  \\
          & $\rightarrow$ Bidirectional GRU: 128 cells \\
    \midrule          
     \cellv{Speaker}{embedding} & 128-D\\
    \midrule
          
     Attention type & 5-component GMM attention w/ softplus \cite{graves2013generating} \\
     Attention RNN & LSTM-256-Zoneout(0.1) $\rightarrow$ FC-128-tanh\\
     DecoderRNN & 2-layer residual-LSTM-256-zoneout(0.1) \\
             & $\rightarrow$ FC-128-Linear\\
    \midrule               
    \rowheader{3}{Mel-spectrogram decoder targets}
        & FFT size: 2048 \\
        & Frame hop: 300 \\
        & Frame length: 1200 \\  
        & Mel bands: 128 \\
    \midrule                
     \cellv{Reduction}{factor} & 2 \\
    \midrule    
     Optimizer & ADAM w/ learning rate $10^{-3}$, batch size 256 \\
    \bottomrule
    \end{tabular}
    \caption{Hyperparameters for the baseline multi-speaker Tacotron system used in this paper.}
    \label{appendix:table:tacotron_hparams}
\end{table}

The attention module uses a single LSTM layer with 256 units and zoneout of 0.1 followed by an MLP with 128 tanh hidden units to compute parameters for the monotonic 5-component GMM attention window. We use GMMv2b attention mechanism described in \cite{battenberg2020location}. Instead of using the exponential function to compute the shift and scale parameters of the GMM components as in \cite{graves2013generating}, GMMv2vb uses the softplus function, and also adds initial bias to these parameters, which we found leads to faster alignment and more stable optimization. The attention weights predicted by the attention network are used to compute a weighted sum of output of the text encoder, producing a context vector. The context vector is concatenated with the output of the attention LSTM layer before being passed to the first decoder LSTM layer. The autoregressive decoder module consists of 2 LSTM layers each with 256 units, zoneout of 0.1, and residual connections between the layers. The $128$-bin mel-spectrogram output is produced using a linear layer on top of the 2 LSTM layers, and we use a reduction factor of 2, meaning we predict two spectrogram frames for each decoder step. The decoder is fed the last frame of its most recent prediction (or the previous ground truth frame during training) and the current context as computed by the attention module. Before being fed to the decoder, the previous prediction is passed through a pre-net with the same same structure used before the text encoder above but its own parameters. We list the full set of Tacotron parameters in Table~\ref{appendix:table:tacotron_hparams}.

\makeatletter
\setlength{\@fptop}{0pt}
\makeatother
\begin{table}[htp]
    \centering
    \begin{tabular}{p{18mm} p{55mm}}
    \toprule
    Module & Hyperparameters    \\
    \midrule       

    \rowheader{2}{Conditioning stack}
        & Mel-spectrogram input \\
        & $\rightarrow$ 5 layers DilatedConv1D-512 \\
        & Each layer: kernel size = 3, $2^i$ conv dilation for layer $i$ \\
        & Repeat-upsampled to match audio frame rate \\
    \midrule      
    \rowheader{1}{Generator}
        & Non-causal WaveNet \cite{yamamoto2020parallel} \\ 
        & 3 blocks of 10 DilatedConv1D-128 layers \\
        & Each layer: kernel size = 3, $2^i$ conv dilation for layer $i$ \\
        & Activation between layers: gated $tanh$ \\ 
        & Output layers: 2 x dense 128 units w/ ReLU \\
        & Final linear projection to dim 1, no activation \\
    \midrule      
    \rowheader{3}{Multi- period critics} 
        & Waveform input \\
        & $\rightarrow$ Critics with periods (2, 3, 5, 7, 11), params as in \cite{vocoder_kong2020hifigan} \\
    \midrule        
    \rowheader{2}{Spectrogram critics}
        & Spectrogram input (linear or mel) \\
        & $\rightarrow$ Conv2D-32, kernel size [3,4], strides=[1,2]) \\
        & $\rightarrow$ Conv2D-32, kernel size [3,4], strides=[1,2]) \\
        & $\rightarrow$ Conv2D-128, kernel size [3,4], strides=[1,2]) \\
        & $\rightarrow$ Flatten() \\
        & $\rightarrow$ Conv1D-1, kernel size 1 \\
    \bottomrule
    \end{tabular}
    \caption{GAN vocoder hyperparameters.}
    \label{appendix:table:vocoder_gan_hparams}
\end{table}

\paragraph{GAN-based vocoder} Except where noted, all experimental results use a GAN-based vocoder with a non-causal WaveNet generator \cite{yamamoto2020parallel} and both waveform and spectrogram critics. The $128$-bin mel-spectrogram input features match the decoder targets, and are fed to the generator through a 5-layer dilated convolution conditioning stack. The output is upsampled to the audio frame rate using repetition, and concatenated with noise before being input to generator stack. The generator outputs 24kHz waveform audio. 

The vocoder critics comprise 1) multi-period waveform critics, configured exactly as in HiFi-GAN \cite{vocoder_kong2020hifigan}, and 2)
both linear- and mel-spectrogram critics, which take as input only spectrogram features extracted from audio. The spectrogram features match those of the decoder targets (see Table \ref{appendix:table:tacotron_hparams}), except that we use a frame hop of 240 for linear spectrograms.

The generator is trained to minimize the reverse KL divergence to the reference waveform in a hybrid f-GAN training setup \cite{nowozin_f-gan:_2016,poole2016improved,shannon2020properties}.
The critics are trained using Jensen-Shannon divergence with an $\ell_1$ feature matching loss applied at every hidden layer.

We list the full set of GAN vocoder parameters in Table~\ref{appendix:table:vocoder_gan_hparams}. 

\clearpage{}%

\end{document}